\documentclass[nofootinbib,twocolumn,preprintnumbers,10pt]{revtex4}
\usepackage{amsmath,amssymb,graphicx,epsfig,hyperref,breakurl,color}
\usepackage{lipsum}

\begin{document}

\title{Role of decay in the searches for double-charm baryons}

\author{Fu-Sheng Yu}
\affiliation{School of Nuclear Science and Technology, Lanzhou University, Lanzhou 730000,   China}

\maketitle

The double-charm baryons with the components of two heavy charm quarks and one light quark, are in a big family of hadron spectroscopy predicted in the quark model. The studies on them are helpful to understand the hadron structure and the nature of the strong interaction. 
With the efforts of searches for decades, the experimental measurements achieved a breakthrough with the first observation of double-charm baryon of $\Xi_{cc}^{++}(ccu)$ via $\Xi_{cc}^{++}\to\Lambda_{c}^{+}K^{-}\pi^{+}\pi^{+}$ by the LHCb collaboration in 2017 \cite{Aaij:2017ueg}. 
Since then, the LHCb collaboration has been leading the experimental studies for doubly charmed baryons \cite{Aaij:2018gfl,Aaij:2018wzf,Aaij:2019dsx,Aaij:2019jfq,Aaij:2019zxa}, due to their significantly large production in the Large Hadron Collider (LHC). 
A very recent measurement is performed by LHCb to search for $\Xi_{cc}^+(ccd)$ in the final state of $\Lambda_c^+K^-\pi^+$ \cite{Aaij:2019jfq}, though no significant signal is observed. 

The decay properties play an important role in the experimental searches of doubly charmed baryons. 
The lowest-lying ground states of double-charm baryons can only decay weakly. Two aspects of weak decays are essential in the relevant studies: the exclusive decay processes and the total decay widths (or lifetimes). 

To search for a new particle, we have to select a most favorable process.
The first report on the evidence of doubly charmed baryon was given by the SELEX collaboration in the process of $\Xi_{cc}^+\to \Lambda_c^+K^-\pi^+$  in 2002 \cite{Mattson:2002vu}. 
However, this result has never been confirmed by any other experiments such as the FOCUS, BABAR, Belle and LHCb  collaborations. 
Especially at the beginning of the LHC running with the expected largest production of doubly charmed baryons until 2013, no significant evidence was found by LHCb using the same process of $\Xi_{cc}^+\to \Lambda_c^+K^-\pi^+$ as what SELEX did \cite{Aaij:2013voa}.
In the early 2017, benefited by the precise measurement of $\Lambda_c^+$ decays by BESIII\cite{Ablikim:2015flg,Ablikim:2016tze}, the theoretical studies on the decays of doubly charmed baryons pointed out that the best processes for their searches are $\Xi_{cc}^{++}\to \Lambda_c^+K^-\pi^+\pi^+$ and $\Xi_c^+\pi^+$ \cite{Yu:2017zst}. They have the largest branching fractions among all the decay channels of double-charm baryons, with all charged final particles to be easily detected at LHCb.
The process used by SELEX is actually misleading to the following experiments. 
With the above suggestion, it achieved the first observation of $\Xi_{cc}^{++}$ in the final state of $\Lambda_c^+K^-\pi^+\pi^+$  in 2017 \cite{Aaij:2017ueg},   and its confirmation via the process of $\Xi_{cc}^{++}\to \Xi_c^+\pi^+$ in 2018 \cite{Aaij:2018gfl}.
An interesting check is given that using the 2012 data of LHCb, it can still significantly observe the double-charm baryon via $\Xi_{cc}^{++}\to \Lambda_c^+K^-\pi^+\pi^+$ \cite{Aaij:2017ueg}. The discovery of doubly charmed baryon was postponed by around five years due to a misleading process used at the beginning.

After the observation of $\Xi_{cc}^{++}$, it is worthwhile to search for $\Xi_{cc}^+$ which is expected with the same production as that of the former one \cite{Chang:2006eu}. The process of $\Xi_{cc}^+\to \Lambda_c^+K^-\pi^+$ is predicted to be the most favorable one among all the decay modes of $\Xi_{cc}^+$ \cite{Yu:2017zst,Li:2017ndo,Wang:2017mqp}.
With the data of a total integrated luminosity of 9 fb$^{-1}$ collected by LHCb, no significant signal of $\Xi_{cc}^+$ is observed \cite{Aaij:2019jfq}. 
Compared to the observation of $\Xi_{cc}^{++}$ using only 1.7 fb$^{-1}$ \cite{Aaij:2017ueg}, it might indicate the smallness of the lifetime of $\Xi_{cc}^+$.

The lifetime is the other important feature to search for double-charm baryons at the hadron-hadron colliders. 
Firstly, the branching fraction of an exclusive decay is proportional to the lifetime, $\mathcal{B}_i = \Gamma_i\cdot \tau$. 
With the same decay dynamics, i.e. the same partial width, a shorter lifetime leads to a smaller branching fraction, along with less signal events. 
Secondly, at the hadron-hadron colliders such as LHC, a shorter lifetime of a particle implies a shorter distance between its primary and secondary vertices. 
In the case that the production vertex is close to the primary vertex in the hadron-hadron collisions, the backgrounds would be larger making the observation more difficult. 
The feature can be clearly seen in \cite{Aaij:2019jfq}, 
for example that the single-event sensitivities $\alpha$ for the 2018 data in Table 3 are 2.36$\pm$0.34, 1.06$\pm$0.15, 0.68$\pm$0.10 and 0.52$\pm$0.08 in the unit of $10^{-2}$, for the $\Xi_{cc}^+$ lifetime hypotheses of 40, 80, 120 and 160 fs, respectively.
With the relation of $\alpha\propto 1/\varepsilon_{\rm sig}$, the signal selection efficiency of the $\Xi_{cc}^+$  decay is approximately proportional to the lifetime, $\varepsilon_{\rm sig} \propto \tau(\Xi_{cc}^+)$, in the range between 40 fs and 160 fs.
Considering the above two effects, the signal events are proportional to the square of lifetime, $N_{\rm sig}\propto \tau^2$, in the above lifetime range.
Therefore, a particle with a shorter lifetime is more difficult to be observed, and vice versa. 
The lifetime of $\Xi_{cc}^{++}$ is measured as large as $256^{+24}_{-22}\pm14$ fs \cite{Aaij:2018wzf},
while that of $\Xi_{cc}^+$  is predicted in most theoretical studies to be about three or four times smaller than the former one \cite{Karliner:2014gca,Kiselev:2001fw,Chang:2007xa,Onishchenko:2000yp}, or even much smaller to be 45 fs in \cite{Cheng:2018mwu}.

Although no significant signal is observed, a small enhancement is seen in the $\Lambda_c^+K^-\pi^+$ final state of $\Xi_{cc}^+$ with a local significance of 3.1$\sigma$, near the mass of $\Xi_{cc}^{++}$ expected under the isospin symmetry \cite{Aaij:2019jfq}. 
It can be expected that $\Xi_{cc}^+$ could be observed with larger data samples and improved trigger conditions by LHCb Run III.  
Further studies on other double-heavy-flavor baryons are also expected in both experiments and theories in the future.

\end{document}